\newif\ifcamerareadypreview
\camerareadypreviewtrue
\ifcamerareadypreview
  \documentclass[cameraready]{Interspeech}
\else
  \documentclass{Interspeech}
\fi
\usepackage{amsmath,amssymb}
\usepackage{graphicx}
\usepackage{booktabs}
\usepackage{url}
\usepackage{hyperref}

\newcommand\blfootnote[1]{%
  \begingroup\renewcommand\thefootnote{}\footnotetext{#1}%
  \addtocounter{footnote}{-1}\endgroup}

\title{Deterministic Prompting for Speaker-Stable Low-Resource Greek TTS}
\author[affiliation={1},orcid={0009-0007-1679-3323}]
{Georgios}{Syllas}

\author[affiliation={2},orcid={0000-0002-6042-9584}]
{Efthymios}{Georgiou*}

\author[affiliation={1},orcid={0000-0003-4513-5040}]
{Kosmas}{Kritsis}

\author[affiliation={3},orcid={0009-0007-1532-5288}]
{Alexandros}{Potamianos}
\address{%
  $^{1}$Institute for Language and Speech Processing, Athena R.C., Greece\\
  $^{2}$Department of Digital Medicine, University of Bern, Switzerland\\
  $^{3}$School of Electrical and Computer Engineering, National Technical University of Athens, Greece}
\email{georgios.syllas@athenarc.gr, efthymios.georgiou@unibe.ch, kosmas.kritsis@athenarc.gr, potam@central.ntua.gr}
\keywords{low-resource TTS, Greek, multilingual pretraining, prompt conditioning, LoRA}

\begin{document}

\maketitle
\blfootnote{\textsuperscript{$\ast$}Work done while EG at the Institute for Language and Speech Processing, Athena Research Center.}

\begin{abstract}
Modern TTS systems approach human quality for high-resource languages but degrade when clean speech data is scarce. Modern Greek exemplifies this, lacking the curated corpora behind state-of-the-art synthesis. We propose a data curation recipe that transforms audiobook recordings into TTS-ready data via WhisperX alignment and filtering. Then we fine-tune Parler-TTS (880M), a prompt-based multilingual model whose pretraining encodes phonetic priors transferable to Greek. During development, we find that LLM-generated style prompts introduce speaker drift at inference. Replacing them with deterministic prompts resolves this, and a speaker-specific LoRA stage trained on 3.5\,h of single-speaker data anchors identity while updating ${\sim}5\%$ of parameters. Our system achieves WER 10.7\% (2.9 above the ASR floor), MOS-I 4.00 (vs.\ 4.36 human speech), and near-human speaker consistency (MOS-C 4.24 vs.\ 4.30), showing that robust single-speaker Greek TTS is achievable with limited curated data.
\end{abstract}

\section{Introduction}
Neural text-to-speech (TTS) has progressed to near-human naturalness in settings where large, clean, and stylistically consistent corpora are available, typically as long single-speaker recordings segmented into well-behaved clips.  However, the same architectures degrade rapidly when training data are scarce or noisy, often harming prosody, intelligibility, and speaker consistency. This high-resource/low-resource gap is especially pronounced for Modern Greek, where public datasets either provide limited clean single-speaker coverage (e.g., CSS10~\cite{css10}) or larger but heterogeneous multi-speaker collections with transcription and acoustic noise (e.g., Common Voice~\cite{commonvoice2023}).

Greek presents challenges beyond data volume. Its rich inflectional morphology and lexical stress system require reliable prosody modeling, while available corpora exhibit speaker and channel variability, inconsistent segmentation, and imperfect transcriptions. In large community datasets such as Common Voice, many speakers contribute only a few minutes each, and fine-tuning on such fragmented supervision drifts toward a diffuse, speaker-averaged voice rather than a stable identity. To address this, our first contribution is a data curation \emph{recipe} that bridges the gap between ``raw audio'' and ``model-ready training examples''. The pipeline aligns long-form recordings to text (WhisperX~\cite{whisperx2023}), segments them into TTS-suitable clips, and filters aggressively for alignment confidence, duration, and acoustic consistency, transforming both community recordings and audiobooks into standardized examples usable for adaptation. We additionally curate two novel audiobook-derived single-speaker Greek datasets to compensate for the lack of clean, long-duration single-speaker material.

Neural TTS has evolved from spectrogram-based sequence-to-sequence models~\cite{Wang2017,shen2018natural,georgiou2023regotron,li2019neural}, through non-autoregressive~\cite{Ren2019,ren2020fastspeech}, flow-based~\cite{GlowTTS2020}, end-to-end~\cite{Kim2021VITS}, and diffusion-based architectures~\cite{popov2021grad}, to codec language models that reframe synthesis as discrete token prediction over neural audio codes~\cite{valle,huggingface2024parlertts,Lyth2024}, supported by vocoders and neural codecs such as HiFi-GAN~\cite{Kong2020} and DAC~\cite{Kumar}. Among these, prompt-conditioned codec models are particularly attractive for low-resource adaptation, as natural-language descriptions provide explicit style control while multilingual pretraining encodes transferable phonetic priors.

We adopt Parler-TTS~\cite{huggingface2024parlertts,Lyth2024}, an 880M-parameter prompt-conditioned multilingual model, because its multilingual pretraining can transfer phonetic and prosodic priors to low-resource languages, especially phonetically close ones~\cite{saeki2024text,Amalas2024,Gong2024}. Parler-TTS is trained on multiple Indo-European languages, including Spanish, which is phonologically and prosodically similar to Greek~\cite{DAUER198351,ARVANITI200797}. We first fully fine-tune the model on our curated Greek data to adapt its multilingual priors to the target language. However, since high-quality single-speaker data remain limited to a few hours, fully updating hundreds of millions of parameters for speaker specialization risks overfitting. We therefore add a second stage using Low-Rank Adaptation (LoRA)~\cite{Hu2022}, training only on single-speaker data with far fewer trainable parameters~\cite{Li2024PETL}.

In early experiments, we find that LLM-generated style prompts, intended to increase linguistic variety and expressiveness, cause a severe speaker consistency failure: the synthesized voice drifts noticeably across utterances, even when conditioning on the same speaker. Replacing stochastic LLM prompts with deterministic, human-designed style prompts resolves this instability. Combined with the speaker-specific LoRA stage trained on 3.5\,h of curated single-speaker Greek, this produces stable identity and high intelligibility.

In summary, we contribute (i) a reusable data curation pipeline that converts raw Greek audio into problem-ready TTS clips, (ii) a two-stage adaptation recipe (full fine-tuning followed by speaker-specific LoRA) that highlights the failure mode of LLM-generated prompts and the benefit of deterministic prompting, and (iii) evidence that stable single-speaker identity can be anchored with only 3.5\,h of speaker-specific data atop the adapted model.

\section{Data Curation Pipeline}
\textbf{Public corpora and cleaning: }
We use two public Greek corpora.  CSS10~\cite{css10} provides ${\sim}$4\,h of
clean single-speaker female audiobook speech.  Mozilla Common Voice
(Greek)~\cite{commonvoice2023} contains ${\sim}$32\,h across 412 speakers.
Common Voice is filtered using community validation signals and manual inspection
to remove clips with clipping, background music, and transcription
mismatches, yielding ${\sim}$17.5\,h of higher-quality multi-speaker speech from 174 retained speakers. After TTS-specific preprocessing for Parler-TTS (format standardization and
duration trimming), we use ${\sim}$15.5\,h for training.

\noindent\textbf{Audiobook-derived single-speaker data: }
To obtain higher-quality single-speaker Greek speech, we build two
audiobook-derived corpora from publicly available recordings via a
semi-automated pipeline: (1)~source selection prioritizing clear, low-noise
single-speaker material; (2)~forced alignment and utterance segmentation via
WhisperX~\cite{whisperx2023} with GPU acceleration; and (3)~quality filtering
via transcription confidence, SNR scoring, and, for the first corpus, manual
review and correction of ASR errors. We constrain segment durations to
approximately 1.5--10\,s to avoid overly short utterances (which can harm
prosody learning).

One corpus is manually verified (${\sim}$3.5\,h, male speaker) and reserved for
speaker-specific LoRA adaptation.  A larger automatically filtered corpus (${\sim}$7.5\,h, male speaker) was excluded from the final pipeline: residual ASR errors propagated into training and increased hallucinated syllables, underscoring that transcription accuracy outweighs raw volume in low-resource TTS.
Table~\ref{tab:data} summarizes all resources.

The source recordings were obtained from privately licensed sources and
cannot be redistributed. Data-preparation templates are available: {\footnotesize \url{https://github.com/gsyllas/greek-stable-tts/tree/main/scripts/data}} to enable
reproduction for researchers with access to original recordings.

We explored voice-conversion augmentation with Seed-VC~\cite{Liu2024SeedVC} to
convert Common Voice samples to the CSS10 speaker timbre, but conversion
artifacts degraded quality and did not improve performance, so we omit this
path from evaluation.

\begin{table}[t]
\centering
\caption{Greek data resources. Durations are approximate after filtering.
  $^\dagger$After community validation; further TTS-specific cleaning yields
  ${\sim}$15.5\,h used in training. Audiobook-2 is excluded from the final system.}
\label{tab:data}
\setlength{\tabcolsep}{3pt}
\small
\begin{tabular}{lccc}
\toprule
\textbf{Dataset} & \textbf{Hours} & \textbf{Spk.} & \textbf{Notes}\\
\midrule
CSS10~\cite{css10} & 4.0 & 1 & Clean, female\\
Common Voice~\cite{commonvoice2023} & 15.5$^\dagger$ & 174 & Filtered\\
Audiobook-1 & 3.5 & 1 & Verified, male\\
Audiobook-2 & 7.5 & 1 & Auto-filtered\\
\bottomrule
\end{tabular}
\end{table}

\section{Models and Adaptation}

\subsection{VITS baseline}

As a low-resource baseline, we fine-tune a pretrained Greek VITS
checkpoint~\cite{Kim2021VITS} on combinations of CSS10, filtered Common Voice and
voice-converted data in various configurations with different data mixtures and
training durations. All configurations exhibit monotonic prosody, timbre
inconsistency, and audible artifacts, with quality insufficient for formal
evaluation, motivating our shift to a multilingual foundation model.

\subsection{Parler-TTS}

Parler-TTS~\cite{huggingface2024parlertts,Lyth2024} frames speech synthesis as
autoregressive generation of discrete audio tokens, jointly conditioned on a
transcript and a natural-language style description.  A frozen Flan-T5
encoder~\cite{Chung2022} processes the style description via cross-attention
layers. A Transformer decoder autoregressively predicts residual vector quantization (RVQ) codebook tokens
across nine codebook levels using a delay-pattern interleaving
scheme~\cite{Copet2023}. A DAC decoder~\cite{Kumar} reconstructs the
final waveform.

We selected the multilingual Parler-TTS checkpoint~\cite{parlertts2023} because
it was pretrained on a diverse language set that includes Spanish, which we
treat as a favorable cross-lingual transfer prior for Greek (not a guarantee).
Its prompt-conditioned design enables explicit style control via text
descriptions, and its modular structure (frozen text encoder, trainable decoder,
frozen audio codec) is well suited to parameter-efficient adaptation.

Full fine-tuning proceeds in two stages: first on Common Voice alone
(${\sim}$15.5\,h) to verify viability, then extended training on the full pool
(${\sim}$23.0\,h: Common Voice + CSS10 + audiobook-1 data). Extended training
improves prosody and naturalness but introduces occasional hallucinated syllables
and introduces timbre drift across utterances, which we mitigate with
deterministic prompts and a speaker-specific LoRA stage.

\noindent\textbf{Deterministic prompt engineering: }
The standard Parler-TTS pipeline generates style descriptions via an LLM from
automatically extracted acoustic attributes (speaking rate, pitch statistics,
SNR, and C50 reverberation). This introduces stochastic wording variation across
training samples and at inference time, destabilizing both optimization and
generation. We replace LLM-generated descriptions with deterministic prompts by
discretizing each scalar attribute into fixed bins (five quantile bins over the
training set) and concatenating the corresponding labels in a fixed order (e.g.,
\textit{``male, slightly low pitch, moderate speed, very clear,
very close-sounding''}). At inference time, we use a single canonical
deterministic prompt (median-bin labels) for all utterances. This removes
prompt-induced variance and reduces hallucinated or repeated syllables, though
speaker identity drift persists without further adaptation.

\noindent\textbf{Speaker stabilization with LoRA: }
A critical failure mode emerges after full fine-tuning on the multi-speaker pool. With many speakers having little per-speaker data, the decoder cannot converge to any
individual voice. Instead it learns a diffuse, speaker-averaged representation
that produces generation-to-generation and utterance-to-utterance timbre drift,
audible even when transcription accuracy is reasonable.  This speaker inconsistency (timbre drift across utterances), rather than lack of
total data, motivates the LoRA stage.

LoRA~\cite{Hu2022} directly targets this failure mode.  By injecting trainable
low-rank projection matrices into the Transformer decoder's attention layers, updating
only ${\sim}$25\,M parameters (${\sim}5\%$ of the 500\,M-parameter decoder), we
anchor the model to a single stable speaker identity without catastrophically
forgetting the multilingual phonetic knowledge acquired during pretraining.
A full re-fine-tuning of the decoder to a single speaker would risk overwriting
those priors. The low-rank constraint acts as an implicit regularizer that
preserves them while specializing the voice.

\begin{table}[t]
\centering
\caption{Compute comparison: full fine-tuning vs.\ LoRA adaptation.}
\label{tab:compute}
\setlength{\tabcolsep}{3pt}
\begin{tabular}{lcc}
\toprule
 & \textbf{Full FT} & \textbf{LoRA}\\
\midrule
Trainable params  & 500\,M (100\%) & 25\,M (5\%)\\
GPU               & A100 40\,GB    & T4 16\,GB\\
Epochs            & 50             & 2\\
Wall time         & ${\sim}$20\,h  & ${\sim}$2\,h\\
\bottomrule
\end{tabular}
\end{table}

\textbf{Implementation details:}
Both stages use the AdamW optimizer with a learning rate of $1\times10^{-4}$.
Full fine-tuning trains the entire 500\,M-parameter decoder for 50 epochs on
the full multi-speaker pool. LoRA then trains for 2 additional epochs on the
3.5\,h single-speaker corpus. Table~\ref{tab:compute} summarizes the resource
comparison.
LoRA adapters are applied to all attention projection matrices
with rank $r\!=\!16$, scaling factor $\alpha\!=\!32$, and dropout $0.05$.
The best checkpoint is selected by lowest validation loss on a held-out 10\%
split (created within each training stage).

Evaluation uses held-out sets: the 50-utterance WER/CER set is drawn from the
Common Voice test partition, and the 20-utterance MCD/SIM-S set is held out from
the audiobook corpus.
At inference, we use greedy decoding.

\section{Evaluation}

\subsection{Objective metrics}

Intelligibility is measured via ASR-based word error rate (WER) and character
error rate (CER) using WhisperX v3 transcriptions on normalized text (lowercased,
punctuation removed) on a 50-utterance held-out Common Voice test set.  We
additionally compute mel-cepstral distortion (MCD) and speaker similarity
(SIM-S; cosine similarity of ECAPA-TDNN~\cite{Desplanques2020} embeddings extracted
via SpeechBrain~\cite{ravanelli2021speechbraingeneralpurposespeechtoolkit}) on 20 utterances from the manually verified single-speaker audiobook corpus.

\subsection{Listening study}

A listening study with 29 native Greek speakers (14 non-expert, 15 expert;
self-reported AI/ML familiarity; headphones required in a quiet environment)
rates samples on 5-point Likert scales for naturalness (MOS-N) and
intelligibility (MOS-I). For MOS-N/MOS-I, each participant evaluated 15 clips
total (3 per system across 5 systems, including ground-truth recordings),
presented in fully randomized order through a custom web interface. We report
MOS-N/MOS-I over respondents who completed all ratings for this task ($n=25$).
No responses were excluded post hoc.  Participation was voluntary with
informed consent; no formal IRB approval was required under our
institutional guidelines for non-clinical perceptual studies.
We use separate, clearly defined tasks for each metric to avoid conflating
naturalness with perceived quality~\cite{chiang2023why}. Vocal consistency (MOS-C) is assessed in a separate comparison task by
  presenting two utterances from the same source and asking participants to rate
  their vocal similarity. This task includes two TTS systems
  (Det.\,+\,LoRA and LLM\,+\,LoRA) and a human reference condition formed by
  pairs of recordings from the manually verified single-speaker audiobook corpus.
  We report results over respondents who completed all comparisons ($n=27$).
\begin{figure*}[!t]
  \centering
  \includegraphics[width=0.98\textwidth]{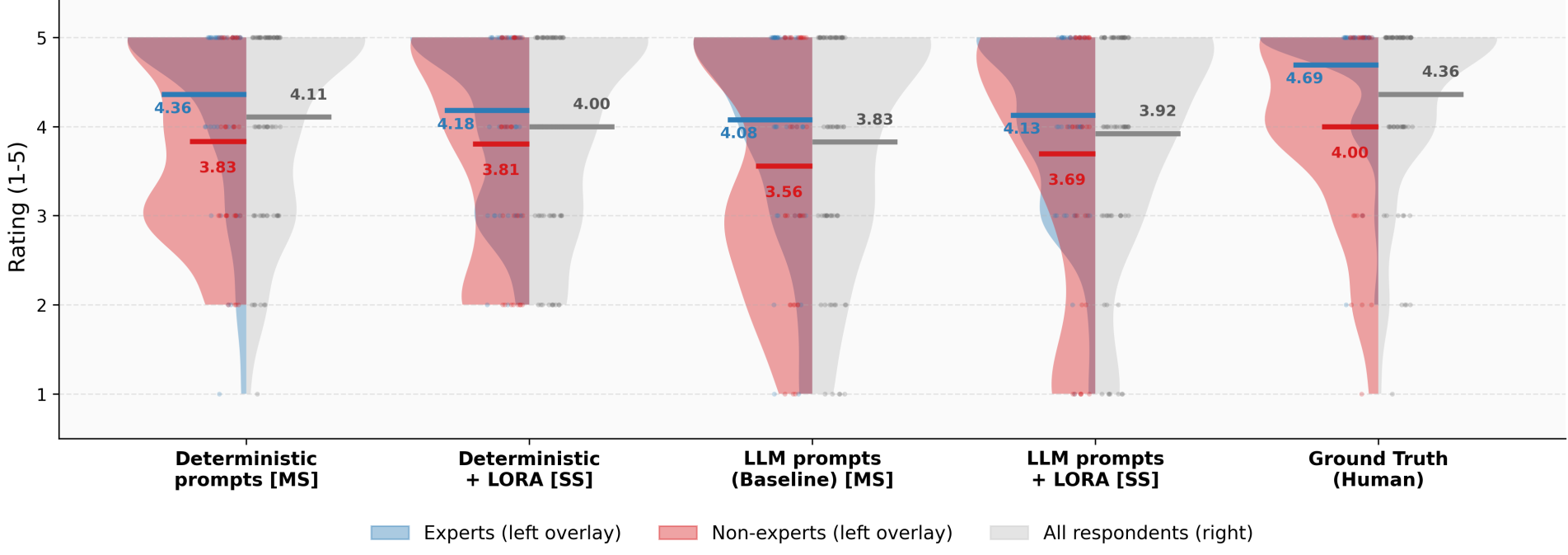}
  
  \caption{Intelligibility (MOS-I) distributions with expert/non-expert overlays
    for all five systems. The Deterministic and LLM (Baseline) systems use
    multi-speaker fine-tuning only, +\,LoRA systems add single-speaker
    adaptation, Ground Truth is reference audio.}
  \label{fig:violin_q1b}
\end{figure*}
\section{Results}
\subsection{Objective intelligibility and speaker similarity}
Table~\ref{tab:objective} reports results for all Parler-TTS configurations.
Without LoRA, LLM-prompted descriptions achieve lower WER than deterministic
descriptions (15.2\% vs.\ 18.8\%); however, once LoRA is applied, the deterministic
configuration substantially outperforms LLM prompts (WER 10.7\% vs.\ 21.1\%).
This synergy arises because LoRA's speaker specialization benefits from the
reduced conditioning variance of deterministic prompts: the model receives a
consistent, predictable style signal that aligns with the stable speaker identity
locked in by LoRA.  The best configuration (Det.\,+\,LoRA) achieves WER\,=\,10.7\%,
only 2.9\,pp above the ASR floor on human audio (7.8\%). However, the absolute speaker-similarity scores remain modest (SIM-S ${\approx}$ 0.60) and MCD does not improve for the best intelligibility configuration, indicating that target-speaker matching and fine-grained spectral fidelity remain limited in this low-resource setting. For this reason, our subjective speaker evaluation focuses on \textit{intra-model voice consistency} (MOS-C), i.e., whether each system maintains a stable voice across utterances, rather than strict similarity to a fixed reference speaker.

Manual inspection of high-WER utterances reveals three recurring failure
modes: \textit{lexical-stress errors}, where the model shifts stress to
the wrong syllable in polysyllabic words, producing phonetically plausible
but semantically incorrect output, \textit{hallucinated syllables}, i.e.\
inserted or repeated sub-word units that inflate WER while CER remains low;
and \textit{punctuation--prosody mismatch}, where sentence-final
intonation does not align with the punctuation mark (e.g.\ declarative
contour on a question).  LoRA adaptation sharply reduces the last two, while the first persists at a low rate across all configurations.
\begin{table}[t]
\centering
\caption{Objective metrics. Setup abbreviations: MS = multi-speaker full
  fine-tuning; SS = multi-speaker full fine-tuning followed by
  single-speaker adaptation. MCD and SIM-S are computed only for LoRA
  configurations. Bold marks the best value per column (lower is better for
  WER/CER/MCD, higher for SIM-S).}
\label{tab:objective}
\setlength{\tabcolsep}{3pt}
\small
\begin{tabular}{llcccc}
\toprule
\textbf{System} & \textbf{Spk.} & \textbf{WER}$\downarrow$ & \textbf{CER}$\downarrow$ & \textbf{MCD}$\downarrow$ & \textbf{SIM-S}$\uparrow$\\
\midrule
Ground Truth     & Ref.    & 7.8\%  & 2.3\%  & --   & --     \\
Parler-TTS (LLM) & MS      & 15.2\% & 6.2\%  & --   & --     \\
Parler-TTS (Det.)& MS      & 18.8\% & 8.0\%  & --   & --     \\
LLM\,+\,LoRA     & SS      & 21.1\% & 7.6\%  & \textbf{8.37} & 0.60\\
Det.\,+\,LoRA    & SS      & \textbf{10.7\%} & \textbf{3.7\%} &  8.98 & \textbf{0.61} \\
\bottomrule
\end{tabular}
\end{table}
\subsection{Subjective quality and consistency}
As seen in Table~\ref{tab:subjective_main}, deterministic prompts without LoRA achieve the highest naturalness and intelligibility MOS among
synthetic systems (MOS-N\,=\,3.76, MOS-I\,=\,4.11), while Table~\ref{tab:objective} shows that the objective WER/CER benefit emerges after LoRA adaptation. The synthetic systems
also achieve slightly higher mean naturalness than the sampled ground-truth
clips (MOS-N\,=\,3.47 for ground truth), which are drawn from Common Voice
recordings with variable microphone quality and ambient noise. The Parler-TTS
backbone, pretrained on tens of thousands of hours of high-quality studio audio,
produces cleaner waveforms.  We attribute this gap to
recording conditions rather than genuine superiority over human speech. The Det.\,+\,LoRA configuration yields near-human
speaker consistency (MOS-C\,=\,4.24 vs.\ 4.30 for a human speaker, a small
absolute difference) while maintaining competitive naturalness and
intelligibility. In contrast, LLM\,+\,LoRA scores substantially lower on MOS-C
(3.56), supporting the conclusion that deterministic prompting is important for
stable speaker identity after speaker-specific adaptation.
To interpret these trends under an ordinal repeated-measures design, we compute
per-listener mean scores per system and apply Friedman tests with pairwise
Wilcoxon signed-rank post-hoc tests using Holm correction.
For naturalness, we find no evidence of differences across systems
($\chi^2=3.83$, $p=0.43$), so apparent mean gaps should be treated as
inconclusive. For intelligibility, the omnibus test is significant
($\chi^2=10.26$, $p=0.036$), indicating that at least one system differs. The
only Holm-significant pairwise result is that the LLM baseline is rated lower
than human recordings ($p_{\text{adj}}=0.025$) with a large effect size
($r=-0.735$; rank-biserial correlation). The remaining synthetic systems are not
significantly different from human recordings after Holm correction, which is
consistent with (but does not prove) human-level intelligibility; with $n=25$
and conservative family-wise error control, small differences may go undetected.
For voice consistency (MOS-C), the omnibus test is not significant
($\chi^2=3.30$, $p=0.19$), but both comparisons against LLM\,+\,LoRA are marginal
after correction ($p_{\text{adj}}=0.076$), suggesting that LLM\,+\,LoRA yields less stable speaker identity than Det.\,+\,LoRA in this low-resource setting.
\noindent\textbf{Exploratory subgroup analysis:} as a secondary descriptive analysis, we split listeners by self-reported
AI/ML/TTS familiarity into non-experts ($n=12$) and experts
($n=13$). Experts assign higher intelligibility scores and exhibit an omnibus
difference across systems for intelligibility (Friedman $\chi^2=10.76$,
$p=0.029$), whereas non-experts do not ($\chi^2=2.12$, $p=0.713$). Within the
expert subgroup, the LLM baseline vs.\ human comparison is marginal after Holm
correction ($p_{\text{adj}}=0.066$, $r=-0.885$). Fig.~\ref{fig:violin_q1b}
visualizes intelligibility distributions by subgroup for all five systems,
highlighting that experts use the upper end of the scale differently from
non-experts across the full comparison set. Given the small subgroup sizes and
multiple comparisons, we treat these findings as descriptive only~\cite{chiang2023why}.

\begin{table}[t]
\centering
\caption{Subjective scores (mean\,$\pm$\,std). Speaker labels: MS = multi-speaker
  model; SS = single-speaker adapted. MOS-C uses paired samples from the same
  system. Bold marks best synthetic score per column.}
\label{tab:subjective_main}
\setlength{\tabcolsep}{3pt}
\footnotesize
\begin{tabular}{llccc}
\toprule
\textbf{System} & \textbf{Spk.} & \textbf{MOS-N} & \textbf{MOS-I} & \textbf{MOS-C}\\
\midrule
Ground truth      & Ref. & 3.47\,$\pm$\,1.26 & 4.36\,$\pm$\,0.94 & \textbf{4.30}\,$\pm$\,0.79\\
Parler-TTS (Det.) & MS   & \textbf{3.76}\,$\pm$\,1.14 & \textbf{4.11}\,$\pm$\,1.07 & --\\
Parler-TTS (LLM)  & MS   & 3.49\,$\pm$\,1.18 & 3.83\,$\pm$\,1.29 & --\\
LLM\,+\,LoRA      & SS   & 3.60\,$\pm$\,1.24 & 3.92\,$\pm$\,1.27 & 3.56\,$\pm$\,1.19\\
Det.\,+\,LoRA     & SS   & 3.68\,$\pm$\,0.90 & 4.00\,$\pm$\,1.09 & 4.24\,$\pm$\,0.78\\
\bottomrule
\end{tabular}
\end{table}
\section{Discussion}
\textbf{Multilingual pretraining} provided cross-lingual phonetic priors that make Greek adaptation feasible under limited data, where a Greek-only baseline did not reach sufficient quality for formal evaluation. \textbf{Deterministic prompts} reduced conditioning variance and yielded the highest mean MOS among synthetic systems, especially when paired with LoRA. \textbf{Speaker-specific LoRA} counteracted speaker averaging and yielded high intra-system voice consistency (MOS-C\,=\,4.24), with trends favoring Det.\,+\,LoRA over LLM\,+\,LoRA ($p_{\text{adj}}=0.076$).
Our experiments are limited to Modern Greek and one male LoRA speaker in a reading style, so multilingual replication and broader speaker/style coverage are future work. We also leave a
causal decomposition of LLM-induced drift, wording variation, semantic mismatch, training-distribution mismatch, and prompt-generation quality, to future work. Most synthetic-vs-synthetic differences are not significant after Holm correction, and non-significant differences from human recordings should not be interpreted as equivalence without an explicit non-
inferiority margin. ASR-based WER may also misestimate perceptual error rates for morphologically complex Greek.

In summary, multilingual transfer, deterministic prompt conditioning, and
speaker-specific LoRA provided a practical recipe for Greek TTS under severe data
and compute constraints.  The best configuration approaches the ASR floor on
human recordings (WER\,=\,10.7\% vs.\ 7.8\%) and achieves near-human
intra-system speaker consistency (MOS-C\,=\,4.24 vs.\ 4.30) despite limited
data and compute.
High-quality single-speaker TTS carries inherent misuse risks, including
voice impersonation and deepfake generation.  Our models are released
strictly for research purposes; we encourage the community to pair such
systems with speaker-consent verification and synthetic-speech watermarking.
Code, model weights, and data-processing scripts are released at: {\footnotesize \url{https://gsyllas.github.io/greek-stable-tts/}}

\section{Acknowledgments}

This work received partial funding from the European High-Performance Computing Joint Undertaking (JU) under Grant Agreement No. 101234269 for the Pharos AI Factory project, as well as from the Greek Ministry of Digital Governance and Artificial Intelligence. We gratefully acknowledge the EuroHPC Joint Undertaking for awarding this project access to the EuroHPC supercomputer LEONARDO, hosted by CINECA (Italy) and the LEONARDO consortium through a EuroHPC Development Access call (Project No. EUHPC-D29-081).

\section{Generative AI Use Disclosure}
The authors used a large language model to assist with language editing.
All technical content and experimental results are based on the authors' work.

\bibliographystyle{IEEEtran}
\bibliography{myBib}

\end{document}